\documentclass{article}

\usepackage{arxiv}

\usepackage[utf8]{inputenc} % allow utf-8 input
\usepackage[T1]{fontenc}    % use 8-bit T1 fonts
\usepackage{hyperref}       % hyperlinks
\usepackage{url}            % simple URL typesetting
\usepackage{booktabs}       % professional-quality tables
\usepackage{amsfonts}       % blackboard math symbols
\usepackage{nicefrac}       % compact symbols for 1/2, etc.
\usepackage{microtype}      % microtypography
\usepackage{lipsum}		% Can be removed after putting your text content

\usepackage{amsmath}
\usepackage{amssymb}
\usepackage{multicol}
\usepackage{graphicx}
\usepackage{caption}

\title{Efficient Mixed Dimension Embeddings for Matrix Factorization}

\author{
  Dmitrii Beloborodov \\
  \texttt{dmitribeloborodov@yandex.ru} \\
  \And
  Andrei Zimovnov \\
  \texttt{zimovnov@gmail.com} \\
  \And
  Petr Molodyk \\
  \texttt{pamolodyk@yandex-team.ru} \\
  \And
  Dmitrii Kirillov \\ %$^{1,2}$
  \texttt{wosadeh@yandex-team.ru} \\ 
  \\
  \emph{Yandex, Moscow, Russia} \\
}

\begin{document}
\maketitle

\bigskip

\begin{abstract}

Despite the prominence of neural network approaches in the field of recommender systems, simple methods such as matrix factorization with quadratic loss are still used in industry for several reasons. These models can be trained with alternating least squares, which makes them easy to implement in a massively parallel manner, thus making it possible to utilize billions of events from real-world datasets. Large-scale recommender systems need to account for severe popularity skew in the distributions of users and items, so a lot of research is focused on implementing sparse, mixed dimension or shared embeddings to reduce both the number of parameters and overfitting on rare users and items. In this paper we propose two matrix factorization models with mixed dimension embeddings, which can be optimized in a massively parallel fashion using the alternating least squares approach.

\end{abstract}

% keywords can be removed

\section{Introduction}

Recommender systems are very important to the industry, as a lot of online services are built around them, including content platforms, marketplaces, social networks and advertisement providers. Industrial applications are also often associated with massive amounts of data, collected from millions of users. To efficiently utilize big data in a recommender system, special algorithms are required, that can be implemented in parallel in a multi-machine setup, e.g. in the MapReduce \cite{dean2008mapreduce} paradigm.

An example of such algorithms are matrix factorizations, which still remain competitive in comparison with more complex methods \cite{rendle2021revisiting,rendle2019difficulty}. In the matrix factorization approach, a large sparse matrix that represents user-item interactions is factorized into two or more dense matrices that contain vector representations for users and items, called embeddings.

Matrix factorization models are appealing for big data applications, because they can be optimized in an alternating least squares manner (ALS) \cite{bell2007scalable,hu2008collaborative}: when item embeddings are fixed, optimal user embeddings can be computed in closed form, and vice versa. ALS has fast convergence and requires only a handful of iterations to achieve high performance. Also, ALS scales linearly with the number of users, items and interactions, and can be efficiently parallelized \cite{mehta2021alx,tan2018matrix,tan2016faster,low2012framework}. However, not all matrix factorization models can be solved with ALS, because not all models allow a solution in closed form. ALS is mostly used for matrix factorizations with quadratic loss.

Besides the large amounts of data, real-world datasets for recommendation are also known for a strong skew in the distribution of the number of interactions (popularity) for both users and items. The difference in the number of interactions for the most popular and the least popular items can be several orders of magnitude, which means that representations with variable capacity might be required.

The challenges of real-world data motivate researchers to explore sparse, mixed dimension or shared representations for users and items, to reduce the number of parameters, improve quality, or both. We review the existing approaches in section \ref{related}. One of the ways to account for popularity skew is to introduce mixed dimension embeddings, where the embedding dimension increases with user or item popularity.

In this paper we propose two matrix factorization models that incorporate mixed dimension embeddings and can be optimized with alternating least squares. The proposed algorithms allow better complexity-quality trade-off than the baseline matrix factorization and can be easily implemented in a parallel manner. We analyse their complexity and compare them on two well-known datasets for movie recommendation.

In section \ref{background} we provide the preliminary knowledge on collaborative filtering, matrix factorizations and alternating least squares. In subsections \ref{sec_zero}, \ref{sec_proj} we describe our proposed methods. We analyse their complexity in subsection \ref{complexity} and discuss ways to assign different embedding dimensions in subsection \ref{scheme}. In section \ref{exp} we describe our experimental setup and provide our results.

We will publicly release our code when our paper is published.

\section{Related Works} \label{related}

Methods for efficient user and item representation can be grouped by the general structure of the representation.

Some approaches use dense, but shared representations. Indices for embedding lookup are determined by hashing or similar techniques, for example with complementary partitions \cite{shi2020compositional}, double hashing \cite{zhang2020model}, cyclic arrays \cite{desai2022random} or binary codes \cite{yan2021binary}. These methods don't take the popularity distribution into account, so usually they reduce the number of parameters at the expense of model quality.

Method \cite{kang2020learning} represents embeddings as a set of cluster centroids to compress existing embedding matrices and trains them end-to-end using a straight-through estimator.

Many approaches simply use dense embeddings with different dimensions \cite{joglekar2020neural,ginart2021mixed,zhaok2021autoemb,zhao2020memory}, sometimes called mixed dimension embeddings. 
Embeddings are then cast to a common dimension using trained projection matrices and plugged into the main method. In \cite{zhao2020memory}, an alternative way to cast embeddings to a common space is explored: embeddings are padded with zeros up to the maximum dimension. This approach yielded worse performance than the version with projection matrices. Individual embedding dimensions are either assigned heuristically \cite{ginart2021mixed}, or optimized with neural architecture search \cite{zhaok2021autoemb,zhao2020memory} or reinforcement learning \cite{joglekar2020neural}.

Similar ideas are also used in natural language processing: \cite{baevski2018adaptive} use mixed dimension embeddings with projection matrices to account for different word frequencies, and \cite{chen2018groupreduce} perform popularity-aware block-wise decomposition of the embedding matrix with varying dimensions.

Some methods make weak assumptions about the embedding structure and directly optimize sparse embeddings. Sparsity is modelled with soft masks, which are optimized either with a gradient descent in an end-to-end fashion \cite{liu2021learnable} or with neural architecture search \cite{cheng2020differentiable}.

Some of the methods above require partitioning users and items by popularity to assign fixed dimensions or dimension candidates for automatic search. We explore partitioning in greater detail in subsection \ref{scheme}.

Regardless of the representation structure and dimension selection method, most of the methods above are trained with variations of gradient descent. To the best of our knowledge, mixed dimension embeddings were not explored before in the context of models trained with alternating least squares.

\section{Background} \label{background}

\subsection{Collaborative Filtering}
% TODO links
The goal of a recommender service is to recommend a set of items (e.g. movies) to each user in order to maximize long-term user engagement. Generally speaking, items and users can be described with arbitrarily complex objects with many properties like categories, demographics or item content.

The collaborative filtering approach, however, allows us to abstract from the complexity of the real world and represent users and items as just elements of a set, each having a unique numerical identifier. All the information we have about the users and items is contained within user-item interactions, induced by some sort of user feedback.

There are two distinct settings for the collaborative filtering approach. The explicit feedback setting assumes that users explicitly articulate their preferences for presented items, for example, by rating them. In case of implicit feedback, we do not know exact user preferences for the presented items, we only know some characteristic of user's engagement with items, like the fact that the user consumed the item. 
In both cases, only for a small number of user-item pairs the feedback is known, and the goal is to estimate the feedback for other pairs.
In this study we only focus on the explicit feedback setting, leaving the implicit case for future work.

\subsection{Matrix Factorization}

The matrix factorization approach to collaborative filtering represents users and items as vectors with a fixed dimension $d$, called embeddings. Each user $u$ is associated with a vector $x_u \in \mathbb{R}^d$, and each item $i$ is associated with a vector $y_i \in \mathbb{R}^d$. The rating $r_{ui}$ for a user-item pair $(u,i)$ is estimated by computing the dot product between the corresponding embeddings: $\widehat{r_{ui}} = x_u^T y_i$.

User and item embeddings are model parameters which are optimized during model training. In the case of a quadratic loss function, optimal embedding values are found by minimizing the expression in equation \ref{als}.

\begin{equation}
    \label{als}
    L(X, Y) = 
    \sum_{(u,i) \in \mathbb{P}} \left(x_u^T y_i - r_{ui} \right)^2
    + \lambda \sum_{i \in \mathbb{I}} ||y_i||^2
    + \lambda \sum_{u \in \mathbb{U}} ||x_u||^2
    \rightarrow \min_{X, Y}
\end{equation}

Where $\mathbb{P}$ is the set of all user-item pairs where the true ratings $r_{ui}$ are known, $\mathbb{I}$ is the set of all items, $\mathbb{U}$ is the set of all users, and $\lambda > 0$ is a regularization coefficient. All embeddings are denoted as $X = \{x_u | u \in \mathbb{U}\}, Y = \{y_i | i \in \mathbb{I}\}$.

This way we effectively factorize a sparse rating matrix of size $|\mathbb{U}| \times |\mathbb{I}|$, containing $r_{ui}$ at positions $(u,i) \in \mathbb{P}$ and unknown values at other positions, into two dense embedding matrices of sizes $|\mathbb{U}| \times d$ for users and $|\mathbb{I}| \times d$ for items.

Other choices for the loss function are possible, for example logloss \cite{johnson2014logistic}, or pairwise loss functions \cite{rendle2012bpr}. However, with a quadratic loss function, optimization can be performed efficiently in the alternating least squares manner, proposed in \cite{bell2007scalable}.

To derive alternating least squares method, let's write only the terms dependent on one user embedding $x_u$ in equation \ref{als_one} and minimize it over $x_u$.

\begin{equation}
    \label{als_one}
    L(x_u, Y) = 
    \sum_{\mathbb{P}(u)} \left(x_u^T y_i - r_{ui} \right)^2
    + \lambda ||x_u||^2
    \rightarrow \min_{x_u}
\end{equation}

Where $\mathbb{P}(u) = \{i \in \mathbb{I}: (u, i) \in \mathbb{P}\}$ is the set of all items that the user $u$ rated.

The gradient of the loss from equation \ref{als_one} over $x_u$ can be found in equation \ref{als_der}.

\begin{equation}
    \label{als_der}
    {L'}_{x_u} = 
    2 \sum_{\mathbb{P}(u)}
    y_i \left(y_i^T x_u - r_{ui} \right)
    + 2 \lambda  x_u
    =
    2 \left( \sum_{\mathbb{P}(u)}
    y_i y_i^T \right) x_u 
    -
    2 \sum_{\mathbb{P}(u)}
    y_i r_{ui}
    +
    2 \lambda  x_u
    = 2 (Y_u^T Y_u x_u - Y_u^T r_u + \lambda x_u) 
\end{equation}

Where $Y_u \in \mathbb{R}^{|\mathbb{P}(u)| \times d}$ is a matrix, which rows are the embeddings of items in $\mathbb{P}(u)$, and $r_u \in \mathbb{R}^{|\mathbb{P}(u)|}$ is a vector of ratings of user $u$ for these items.

Solving for ${L'}_{x_u} = 0$, we get equation \ref{als_step} for the optimal $x_u$, given that all item embeddings are fixed.

\begin{equation}
    \label{als_step}
    x_u = 
    (Y_u^T Y_u + \lambda I) ^ {-1} Y_u^T r_u
\end{equation}

Here $I$ is the identity matrix of the appropriate size.

Note that the loss function \ref{als} is symmetrical w.r.t. user and item embeddings, so the optimal values for $y_i$ will be symmetric to equation \ref{als_step}. Also when all item embeddings are fixed, updates for different user embeddings are independent (and vice versa) and can be done in parallel, for example, by using the MapReduce framework \cite{dean2008mapreduce}. This property allows us to perform matrix factorization in a massively parallel manner for datasets containing billions of interactions.

After random initialization, embeddings $X,Y$ are computed by updating all $x_u$ and then all $y_i$ using equations of form \ref{als_step}. This process is repeated for a fixed number of iterations, or until convergence, hence the name "alternating least squares".

\section{Mixed Dimension Embeddings for Alternating Least Squares} \label{mde_sec}

Note that equation \ref{als_one} resembles the loss function for linear regression with $x_u$ as regression coefficients, $Y_u$ as the feature matrix and $r_u$ as the targets. This regression problem has $d$ features and $|\mathbb{P}(u)|$ observations. 

The number of observations for different users $u$ may vary significantly in real world datasets, spanning values from less than 10 to hundreds of thousands, especially if no dataset filtering is used. And as large-scale datasets require rather high values of the embedding dimension $d$ to fit properly, a problem arises. What happens to the users with small $|\mathbb{P}(u)|$, compared to the dimension $d$? In terms of linear regression it would mean that the number of features is greater than the number of observations, which would lead to severe overfitting. 

It makes sense to reduce the feature space for users and items with a small number of ratings, which basically means using smaller embeddings. But then we have to deal with mixed dimension embeddings. In this paper we discuss two ways to correctly define mixed dimension embeddings and compare these approaches.

We denote individual user embedding sizes as $d_u \in \{1,\dots, d\}$, and item sizes as $t_i \in \{1,\dots, d\}$. We will discuss ways to choose these values later in subsection \ref{scheme}, for now we can assume that the embedding size increases with the increasing number of ratings for the user or item.

The first approach is to simply pad shorter embeddings with zeros up to the maximum dimension $d$ and then proceed as before. This way is straightforward and allows to solve for $x_u$ and $y_i$ with little modification to equation \ref{als_step}.

The second approach assumes that embeddings of size $k < d$ are projected into a common space $\mathbb{R}^d$ with a matrix of size $d \times k$. This method still allows to use alternating least squares technique. This approach, though more complex, seems to yield better performance.

\subsection{Zero Padded Mixed Dimension Embeddings} \label{sec_zero}

In this subsection we assume that user and item embeddings $x_u$ and $y_i$ still have the same dimension $d$, but the last $d-d_u$ (for users) or $d-t_i$ (for items) components of the embeddings are always equal to zero.

In this case, the loss function is still the same as per equation \ref{als}, but constraints for zeroed components are added in equation \ref{als_zero}.

\begin{equation}
    \label{als_zero}
    \begin{cases}
    L_{zero}(X, Y) = 
    \sum_{(u,i) \in \mathbb{P}} \left(x_u^T y_i - r_{ui} \right)^2
    + \lambda \sum_{i \in \mathbb{I}} ||y_i||^2
    + \lambda \sum_{u \in \mathbb{U}} ||x_u||^2
    \rightarrow \min_{X, Y}, \\
    s.t. \;\;\; x_{u}[d_u:d] = 0, \;\;\; y_{i}[t_i:d] = 0, \;\;\; \forall u \in \mathbb{U}, \;\;\; \forall i \in \mathbb{I}
    \end{cases}
\end{equation}

Where $d_u$ is the true embedding size for user $u$, $t_i$ is the true embedding size for item $i$, and we use the notation $x[a:b]$ to denote a slice operation, which returns a vector composed of all components of the vector $x$ starting from component $a$ (including) and ending with $b$ (excluding), and component count starts from 0.

Note that $x_u^T y_i = x_u[0:d_u]^T y_i[0:d_u] = x_u[0:t_i]^T y_i[0:t_i]$ because of the imposed constraints: we can compute the dot product using only the first $d_u$ or $t_i$ components, since the rest are equal to zero in at least one of the embeddings. For the same reason, $||x_u||^2 = ||x_u[0:d_u]||^2$. Using these properties, let's write only the terms dependent on $x_u$ in expression \ref{als_zero_one}, similar to equation \ref{als_zero}.

\begin{equation}
    \label{als_zero_one}
    \begin{cases}
    L_{zero}(x_u, Y) = 
    \sum_{\mathbb{P}(u)} \left(x_u[0:d_u]^T y_i[0:d_u] - r_{ui} \right)^2
    + \lambda ||x_u[0:d_u]||^2
    \rightarrow \min_{x_u[0:d_u]} \\
    x_{u}[d_u:d] = 0
    \end{cases}
\end{equation}

This equation is similar to \ref{als_step}, with $x_u[0:d_u]$ instead of $x_u$ and $y_i[0:d_u]$ instead of $y_i$. Solving for $x_u[0:d_u]$ in a similar manner, we get equation \ref{als_zero_step}.

\begin{equation}
    \label{als_zero_step}
    \begin{cases}
    x_u[0:d_u] = 
    (\bar{Y}_u^T \bar{Y}_u + \lambda I) ^ {-1} \bar{Y}_u^T r_u \\
    x_{u}[d_u:d] = 0
    \end{cases}
\end{equation}

Here $\bar{Y}_u$ is a matrix of size $
|\mathbb{P}(u)| \times d_u$, which rows are sliced item embedding vectors $y_i[0:d_u]$ from the set $\mathbb{P}(u)$ of items rated by user $u$, and $r_u$ is still a vector of ratings of user $u$ for these items.

Note that some of the vectors $y_i[0:d_u]$ may contain zeros in a few last components (when $t_i < d_u$), but that does not affect the update formula for $x_u$ in any way. 

All the benefits of equation \ref{als_step} discussed earlier still apply: updates for users and items are symmetric, and when user embeddings are fixed, different item embeddings can be computed independently in parallel (and vice versa).

\subsection{Projected Mixed Dimension Embeddings} \label{sec_proj}

In this subsection we assume that embeddings $x_u$ and $y_i$ actually have different dimensions $d_u$ and $t_i$ respectively. We introduce matrices $A_{d_u} \in \mathbb{R}^{d \times d_u}$ and $B_{t_i} \in \mathbb{R}^{d \times t_i}$ to project the embeddings into a common space $\mathbb{R}^d$. We denote projected embeddings as $\bar{x}_u = A_{d_u} x_u$ and $\bar{y}_i = B_{t_i} y_i$. The rating is predicted as the dot product of projected embeddings $\widehat{r_{ui}} = \bar{x}_u^T \bar{y}_i$. The matrix index denotes a source dimension of the projection, and the target dimension is always $d$. We set the matrices with the maximum source dimension to identity matrices: $A_d = I, B_d = I$, and these matrices are constant, while all other matrices are trained along with the embeddings. There are as many $A_{d_u}$ matrices, as unique embedding dimensions amongst users. The same is true for the item matrices. We denote the sets of all trained matrices as $A = \{A_p | p: \exists u \in \mathbb{U}: p = d_u < d\}$ and $B = \{B_p | p: \exists i \in \mathbb{I}: p = t_i < d\}$.

The objective function for this approach is defined in equation \ref{als_proj}, and we need to minimize it for all embeddings and all trainable projection matrices.

\begin{equation}
    \label{als_proj}
    L_{proj}(X, Y, A, B) = 
    \sum_{(u,i) \in \mathbb{P}} \left(\left(A_{d_u} x_u\right)^T \left(B_{t_i}y_i\right) - r_{ui} \right)^2
    + \lambda \sum_{i \in \mathbb{I}} ||y_i||^2
    + \lambda \sum_{u \in \mathbb{U}} ||x_u||^2
    + \beta \sum_{M \in A \cup B} ||M||^2_F
    \rightarrow \min_{X, Y, A, B}
\end{equation}

Here $||M||_F$ is the Frobenius norm for matrix $M$, and $\beta$ is another regularization coefficient. This way we regularize all trainable projection matrices.

Expression \ref{als_proj} can also be optimized with alternating least squares: each of the $X, Y, A, B$ sets of parameters can be computed in a closed form, if all three other sets are fixed. We will derive formulae for these updates, starting with the updates for embeddings $X, Y$.

We can write the terms dependent only on one user embedding $x_u$ to get equation \ref{als_proj_one}.

\begin{equation}
    \label{als_proj_one}
    L_{proj}(x_u, Y, A, B) = 
    \sum_{\mathbb{P}(u)} \left(x_u^T A_{d_u}^T B_{t_i} y_i - r_{ui} \right)^2
    + \lambda ||x_u||^2
    \rightarrow \min_{x_u}
\end{equation}

It is easy to see that it is similar to equation \ref{als_one}, but with $y_i$ replaced by $A_{d_u}^T B_{t_i} y_i = A_{d_u}^T \bar{y}_i$. So we can immediately derive a solution for it in expression \ref{als_proj_step}.

\begin{equation}
    \label{als_proj_step}
    x_u = 
    (\widetilde{Y}_u^T \widetilde{Y}_u + \lambda I) ^ {-1} \widetilde{Y}_u^T r_u
\end{equation}

Where $\widetilde{Y}_u$ is a matrix of size $|\mathbb{P}(u)| \times d_u$, which rows are vectors $A_{d_u}^T \bar{y}_i \in \mathbb{R}^{d_u}$ for items from $\mathbb{P}(u)$, and $r_u$ is a vector of ratings of user $u$ for these items. Item embeddings are updated in a similar manner, since the equations are symmetric w.r.t. items and users.

Note that it is a good idea to compute and store all projected item embeddings $B_{t_i} y_i = \bar{y}_i$ before updating user embeddings, to avoid computing them multiple times.

Now, to derive updates for projection matrices, we need a statement \ref{trick}.

\begin{equation}
    \label{trick}
    x^T M y = flatten(M)^T flatten(xy^T), \;\;\; x \in \mathbb{R}^n, y \in \mathbb{R}^m, M \in \mathbb{R}^{n \times m}
\end{equation}

Where $flatten(M)$ is an operation that transforms a matrix $M \in \mathbb{R}^{n \times m}$ into a vector of size $mn$ by concatenating all rows of the matrix.

\paragraph{Proof}

\begin{displaymath}
x^T M y = \sum_{i=1}^n \sum_{j=1}^m x_i M_{ij} y_j = \sum_{i=1}^n \sum_{j=1}^m M_{ij} (x_i y_j) = \sum_{i=1}^n \sum_{j=1}^m M_{ij} \left( x y^T\right)_{ij}
\end{displaymath}

Now we can replace the double index $(i,j)$ with a single index $k$, which is the index of the elements in the flattened matrix.

%$$  \left\{ reindex (i,j) \rightarrow k \right\} =$$

\begin{displaymath}
x^T M y =\sum_{k=1}^{mn} flatten(M)_k flatten(xy^T)_k = flatten(M)^T flatten(xy^T)
\end{displaymath}

Which gives us statement \ref{trick}. $\blacksquare$

Now let's write only the terms from the loss \ref{als_proj} that depend on one of the matrices $A_p$ in equation \ref{als_proj_one_2}.

\begin{equation}
    \label{als_proj_one_2}
    L_{proj}(X, Y, A_p, B) = 
    \sum_{(u,i) \in \mathbb{P}: d_u=p}
    \left(x_u^T A_p^T B_{t_i} y_i    - r_{ui} \right)^2
    + \beta ||A_p||^2_F
    \rightarrow \min_{A_p}
\end{equation}

Here we sum over all interactions of all the users that were assigned dimension $p$. We can use the definition $\bar{y}_i = B_{t_i} y_i$, as well as statement \ref{trick} to rewrite:

\begin{displaymath}
x_u^T A_p^T B_{t_i} y_i = x_u^T A_p^T \bar{y}_i = \bar{y}_i^T A_p x_u
= flatten(A_p)^T flatten(\bar{y}_i x_u^T)
\end{displaymath}

We also use $||A_p||^2_F = \sum_{i=1}^d \sum_{j=1}^p (A_p)_{ij}^2 = \sum_{k=1}^{pd} flatten(A_p)_k^2 = ||flatten(A_p)||^2$ and finally rewrite equation \ref{als_proj_one_2} as \ref{als_proj_one_3}.

\begin{equation}
    \label{als_proj_one_3}
    L_{proj}(X, Y, A_p, B) = 
    \sum_{(u,i) \in \mathbb{P}: d_u=p}
    \left(flatten(A_p)^T flatten(\bar{y}_i x_u^T)    - r_{ui} \right)^2
    + \beta ||flatten(A_p)||^2
    \rightarrow \min_{A_p}
\end{equation}

Note that expression \ref{als_proj_one_3} is very similar to \ref{als_one} that we solved earlier, with $flatten(A_p)$ instead of $x_u$, $flatten(\bar{y}_i x_u^T)$ instead of $y_i$, $\beta$ instead of $\lambda$ and a sum over a different set of interactions. This expression is also similar to the objective of linear regression, so we can solve it for $flatten(A_p)$ in a similar manner, which yields equation \ref{als_proj_step_2} for the update of $A_p$.

\begin{equation}
    \label{als_proj_step_2}
    flatten(A_p) = 
    (Q_p^T Q_p + \beta I) ^ {-1} Q_p^T r_p
\end{equation}

Where $Q_p$ is a matrix of size $|\{(u,i) \in \mathbb{P} | d_u=p\}| \times pd$, which rows are vectors $flatten(\bar{y}_i x_u^T) \in \mathbb{R}^{pd}$ computed over all user-item pairs that have a user with assigned dimension $p$. Then $r_p$ is a vector of ratings for those pairs. To obtain $A_p$ we just need to revert the $flatten$ operation, which just means slicing a vector $flatten(A_p)$ into $d$ pieces of size $p$ and stacking them as rows to form the matrix $A_p$.
As before, the model is symmetrical w.r.t. items and users, so $B_p$ are computed in a similar manner.

And as before, different $A_p$ can be computed independently, when $X, Y, B$ are fixed. However, there are not many different matrices $A_p$ (just as many as distinct embedding sizes amongst users), so computing updates in parallel over $p$ will not allow us to update $A$ in a massively parallel fashion. To do that, we may need to compute $Q_p^T Q_p \in \mathbb{R}^{pd \times pd}$ and $Q_p^T r_p \in \mathbb{R}^{pd}$ in parallel over users or items, and then apply equation \ref{als_proj_step_2} for each $p$. Here's how we can do that:

\begin{displaymath}
Q_p^T Q_p = \sum_{(u,i) \in \mathbb{P}: d_u=p}
flatten(\bar{y}_i x_u^T) flatten(\bar{y}_i x_u^T)^T
= \sum_{u \in \mathbb{U}: d_u=p} \sum_{i \in \mathbb{P}(u)} flatten(\bar{y}_i x_u^T) flatten(\bar{y}_i x_u^T)^T
\end{displaymath}

\begin{displaymath}
Q_p^T r_p = \sum_{(u,i) \in \mathbb{P}: d_u=p}
r_{ui} flatten(\bar{y}_i x_u^T)
= \sum_{u \in \mathbb{U}: d_u=p} \sum_{i \in \mathbb{P}(u)} r_{ui} flatten(\bar{y}_i x_u^T)
\end{displaymath}

So we can calculate the inner sums in parallel over users, and then sum the results.

Before the update to $A$, it is also a good idea to precompute all $\bar{y}_i = B_{t_i} y_i$ to avoid computing them multiple times.

Finally, a single iteration of training the projected mixed dimension model consists of the following steps:

\begin{enumerate}
    \item Compute all $\bar{x}_u$
    \item Compute all $B_p$ using a formula symmetric to \ref{als_proj_step_2}
    \item Compute all $\bar{y}_i$
    \item Compute all $A_p$ using formula \ref{als_proj_step_2}
    \item Compute all $\bar{x}_u$
    \item Compute all $y_i$ using a formula symmetric to \ref{als_proj_step}
    \item Compute all $\bar{y}_i$
    \item Compute all $x_u$ using formula \ref{als_proj_step}
\end{enumerate}

We update the parameters in order $B, A, Y, X$, to update smaller sets of parameters first, but other orderings are possible too. After a random initialization of all parameters, these steps are repeated for a fixed number of iterations or until convergence.

\subsection{Methods Complexity} \label{complexity}

To express complexity, we use the following notation: $d$ -- maximum embedding dimension, $U = |\mathbb{U}|$ -- total number of users, $I = |\mathbb{I}|$ -- total number of items, $N = |\mathbb{P}|$ -- total number of user-item pairs with known ratings, $N_u = |\mathbb{P}(u)|$ -- number of ratings known for user $u$, $N_p$ -- number of pairs from $\mathbb{P}$ having user assigned dimension $p$, $\delta$ -- number of unique embedding dimensions.

First we derive the complexity of the baseline embedding update \ref{als_step}. Computing $Y_u^T Y_u + \lambda I$ requires $O(d^2 N_u)$ operations, since $Y_u$ has a size of $N_u \times d$. Computing $Y_u^T r_u$ requires $O(dN_u)$ operations, and this term can be ignored. Computing $Y_u^T Y_u + \lambda I$ for all users requires $O(d^2\sum_{u \in \mathbb{U}}N_u) = O(d^2N)$ operations. Updates for items are symmetric, so the same computations for items also require $O(d^2N)$. Then, solving a system of linear equations of size $d$ in equation \ref{als_step} for each user requires at most $O(d^3)$ operations. We need to solve such a system for all users and items, which requires $O(d^3(U+I))$.  The total complexity for one iteration of the baseline is $O(d^3(U+I) + d^2N)$.

The mixed dimension variant with zero padding \ref{als_zero_step} actually requires strictly less computation time than the original matrix factorization, because the update formula is the same, but the matrix $\bar{Y}_u$ has a size of $N_u \times p$, where $p \leq d$ is the effective embedding dimension, and $p < d$ at least for some users and items. However, the asymptotic complexity of one iteration is still the same as for the baseline: $O(d^3(U+I) + d^2N)$.

Now we derive the complexity for the projected embeddings method. Note that before any parameter updates we precompute the projected embeddings $\bar{x}_u = A_{d_u} x_u$ or $\bar{y}_i = B_{t_i} y_i$ for all users or items to avoid computing them multiple times. The complexity of this computation is $O(d^2U)$ and $O(d^2I)$, and these terms can be ignored in favor of terms with higher complexity.

We start with the updates for the embeddings in equation \ref{als_proj_step}. Computing all $A_p^T \bar{y}_i$ for one user $u$ with embedding size $p \leq d$ requires $O(p d N_u)$ operations, which for all users is $O(d^2 N)$. Computing $\widetilde{Y}_u^T \widetilde{Y}_u + \lambda I$ for one user requires $O(p^2N_u)$, which makes it also $O(d^2N)$ for all users. Computing $\widetilde{Y}_u^T r_u$ requires $O(pN)$, and this term can be ignored. This equation is symmetric w.r.t. users and items, so for the items the complexity is also $O(d^2N)$. Solving a system of linear equations of size $p$ takes at most $O(p^3)$ operations for one embedding, and solving these systems for all users and items takes $O(d^3 (U+I))$. So the final complexity of embedding updates is the same as for the baseline: $O(d^3(U+I) + d^2N)$.

Next are the updates for the projection matrices in equation \ref{als_proj_step_2}. Computing $flatten(\bar{y}_i x_u^T) \in \mathbb{R}^{pd}$ over all ratings of all users with dimension $p$ takes $O(pdN_p)$, which for all user projection matrices is $O(d^2N)$. Computing $Q_p^T Q_p + \beta I$ requires $O(p^2d^2 N_p)$, since $Q_p^T$ has a size of $N_p \times pd$. Item projection matrices have symmetric updates, so for all projection matrices it is $O(d^4 N)$, which becomes prohibitive for the case of large dimension $d$.

To mitigate this issue, we compute $Q_p$ using a subsample of $\{(u,i) \in \mathbb{P} | d_u=p\}$: we sample interactions with a probability of $\frac{1}{p^2}$, which reduces the complexity of computing one $Q_p^T Q_p$ to $O(d^2N_p)$, and all of them to $O(d^2N)$. Computation of $Q_p^Tr_p$ is less demanding, so this term can be ignored.

Solving a system of linear equations of size $pd \times pd$ in expression \ref{als_proj_step_2} requires at most $O(p^3d^3)$, which is $O(d^6\delta)$ total for all matrices in $A$ and $B$. 

Finally, the total complexity of the projected embeddings method with subsampling is $O(d^3(U+I) + d^2N + d^6\delta)$. The first two terms are the same as for the baseline, and the last term does not depend on the dataset size, only on hyperparameter choice.

The complexity of all three methods is linear w.r.t. the dataset size, and all updates can be done in a massively parallel fashion, which means that all three methods are applicable to the big data scenario.

\subsection{Mixed Dimension Embeddings Scheme} \label{scheme}

In this subsection we discuss how to assign different embedding dimensions for each user and item. Earlier in this section we argued that users and items with a smaller number of interactions (often referred to as popularity or frequency) require smaller embeddings to avoid overfitting. It makes sense to assign larger embeddings to more popular users and items, but there are many ways to do so.

In many prior works items and users are sorted by popularity and then split into partitions with equal or manually selected sizes \cite{kang2020learning,joglekar2020neural,cheng2020differentiable}. Similar approaches are used in natural language processing to account for word frequencies \cite{baevski2018adaptive,chen2018groupreduce}.

Other works use feature-wise partitioning: each categorical feature gets a different dimension, which is the same for all values of the feature \cite{zhao2020memory,ginart2021mixed}.

Then for each partition the embedding dimension is either heuristically assigned, or automatically discovered during training.

Approach \cite{ginart2021mixed} uses feature-wise partitioning for the CTR prediction task, but for the collaborative filtering task they propose to split users and items by popularity into partitions with equal sums of popularities per partition. Each partition then gets an embedding dimension proportional to the inverse size of the partition to the power of the temperature hyperparameter $\alpha \in [0,1]$. They provide theoretical justification of this approach by analysing the spectral decay of the target matrix under block structure assumption. Though this method is theoretically justified, unfortunately, we failed to reproduce it in our setup.

Some approaches select individual embedding sizes automatically as a part of their method, without any partitioning \cite{zhaok2021autoemb,liu2021learnable}.

Our method does not allow automatic selection of embedding dimensions, so we also utilize an heuristic approach. To avoid any explicit partitioning, we directly compute the embedding dimension from the user's popularity as $d_u = round\left(\frac{ f_u}{\gamma f_{med}}\right)$, where $f_u$ is the user's popularity, $f_{med}$ is the median user popularity, $\gamma$ is a hyperparameter, and the $round()$ function rounds its argument to the closest value from the list of allowed dimensions. We normalize by $f_{med}$ to make the optimal $\gamma$ independent of the absolute values of popularity, which makes it easier to tune the hyperparameters. We treat items in a similar way. This approach lacks theoretical justification, but it is trivial to implement and seems to work well in practice.

\section{Experiments} \label{exp}

\subsection{Datasets}

We test our methods on two well-known datasets for collaborative filtering: Netflix \cite{bennett2007netflix} and MovieLens 25M \cite{harper2015movielens}. Both of these datasets contain users' ratings for a set of movies, ranging from 1 to 5. We binarize the ratings by replacing ratings $\leq 2$ with 0 and ratings $\geq 4$ with 1, and removing other ratings. 

We split each dataset into train, validation and test sets by timestamp, taking $10\%$ of ratings for validation and another $10\%$ for testing. The test set contains the most recent ratings, while the train set contains the oldest ones.

We retain only users and items that have at least 5 ratings in the train set (non-recursively). In the validation and test sets, we retain only users and items that are present in the train set.

\begin{figure}[ht]
  \centering
  \includegraphics[width=0.5\linewidth]{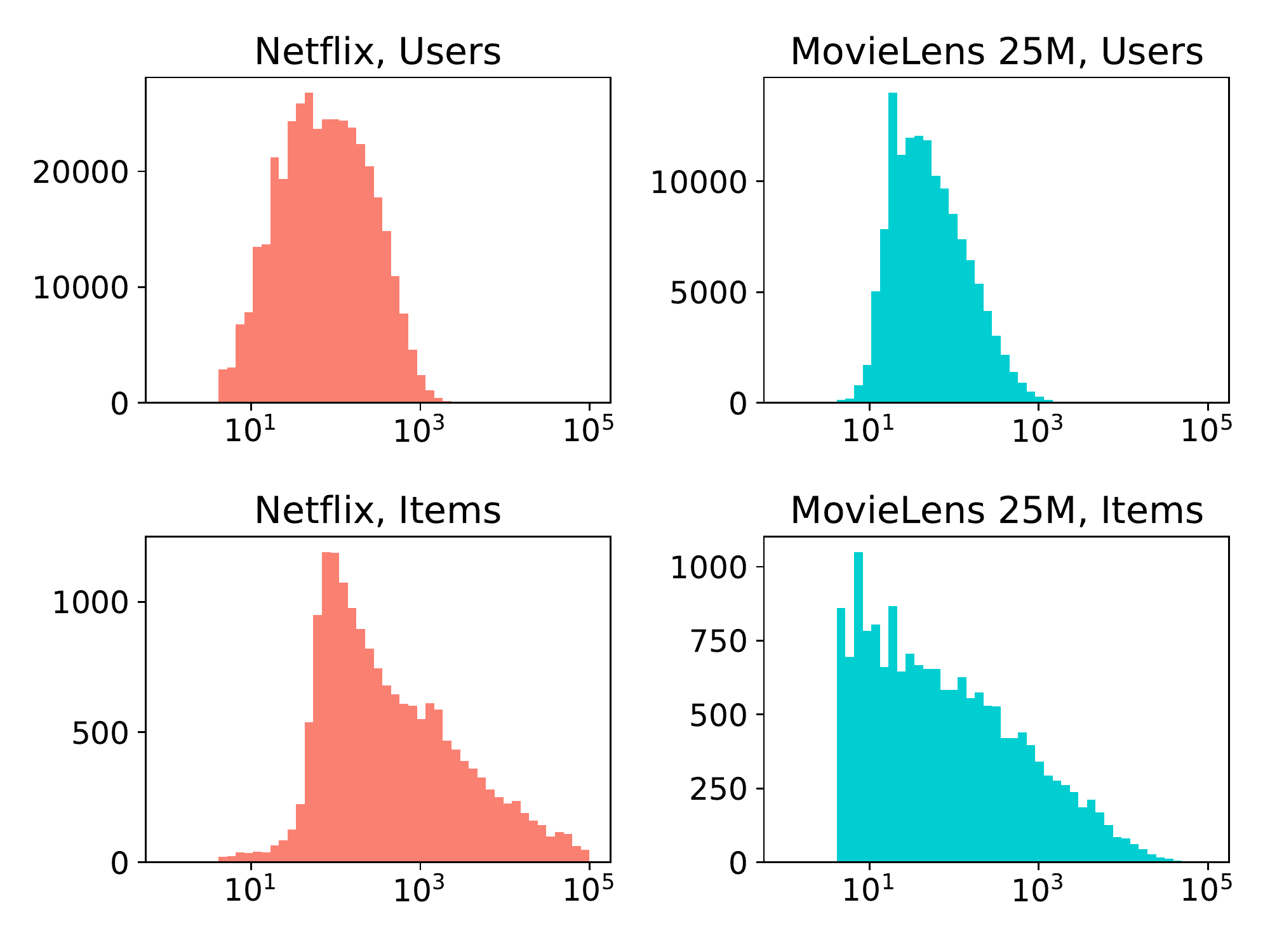}
  \caption{Distributions of user and item popularity in the data in $\log_{10}$ scale.}
  %\Description{Left column correspond to Netflix dataset, right column corresponds to MovieLens dataset. Top row corresponds to user distributions, bottom row to item distributions. All histograms demonstrate heavy-tailed distributions with heavier tails for items.}
  \label{distr}
\end{figure}

The distribution of $\log_{10}$ number of ratings (popularity) for both datasets is presented in figure \ref{distr}. The distributions were computed over the train set after all pre-processing. The skew in popularity is clearly visible, especially for items.

\subsection{Implementation details}

To avoid overfitting, we select the maximum embedding dimension $d=6$ for both datasets. We use dimensions $\{2,4,6\}$ as candidates for the mixed dimension scheme. Individual user and item embedding dimensions are assigned according to subsection \ref{scheme}, proportional to the number of ratings with parameter $\gamma$.

All non-zero embedding components are initialized from the uniform distribution in $[-0.1, 0.1]$. For trained projection matrices, we use the Xavier initialization \cite{glorot2010understanding}. Projection matrices that correspond to users and items with the maximum dimension are constant identity matrices.

We perform 30 iterations of alternating least squares for each training run, evaluate on a hold-out set each 5 iterations and then select the parameters at the best iteration.

\subsection{Evaluation}

To evaluate all methods, we use the ROC AUC score. Results for RMSE are very similar, so we don't include them in the paper.

For each method, we evaluate the performance for models with a different number of parameters to estimate the trade-off between quality and complexity. To vary the number of parameters of the baseline, we change the dimension $d$ in the range $[2,3,4,5,6]$. For the mixed dimension approaches, we fix $d=6$ and vary the hyperparameter $\gamma$ in the range $[0.2, 0.3, 0.5, 1]$.

We use the validation set to select the best iteration during training, and the test set to evaluate the model.
We retrain each model 3 times from scratch to evaluate the variance that is caused by random initialization.

\subsection{Hyperparameters search}

Before evaluating the models, we select the best regularization hyperparameters, using grid search.
To do that, we combine the train and validation sets and then split them randomly into 3 folds for cross-validation. $10\%$ of each train fold is further separated randomly for best iteration selection. Values for the embedding regularization $\lambda$ are selected from the set $\{0.1, 0.3, 1.0, 3.0\}$, values for the projection matrix regularization $\beta$ are selected from $\{300, 1000, 3000, 10000\}$ -- this regularization coefficient requires larger values due to the larger parameter dimensionality.

For each method, we select the best hyperparameters for the largest model and propagate them to smaller models to estimate how each of the methods scale. For the baseline, we select hyperparameters for the model with $d=6$, for the mixed dimension methods we select hyperparameters for the model with $\gamma=0.2$.

\subsection{Experimental results}

We compare matrix factorizations with zero padded and projected embeddings to the baseline matrix factorization with a fixed embedding dimension, all trained with alternating least squares (ALS).

To do that, we plot ROC AUC versus the number of parameters, which includes the sum of all user and item dimensions $d_u$ and $t_i$, as well as the sizes of the projection matrices. Note that the projection matrices are tiny, compared to the embedding matrices, so their contribution is not visible on the plots. 

\begin{figure}[ht]
  \centering
  \includegraphics[width=0.5\linewidth]{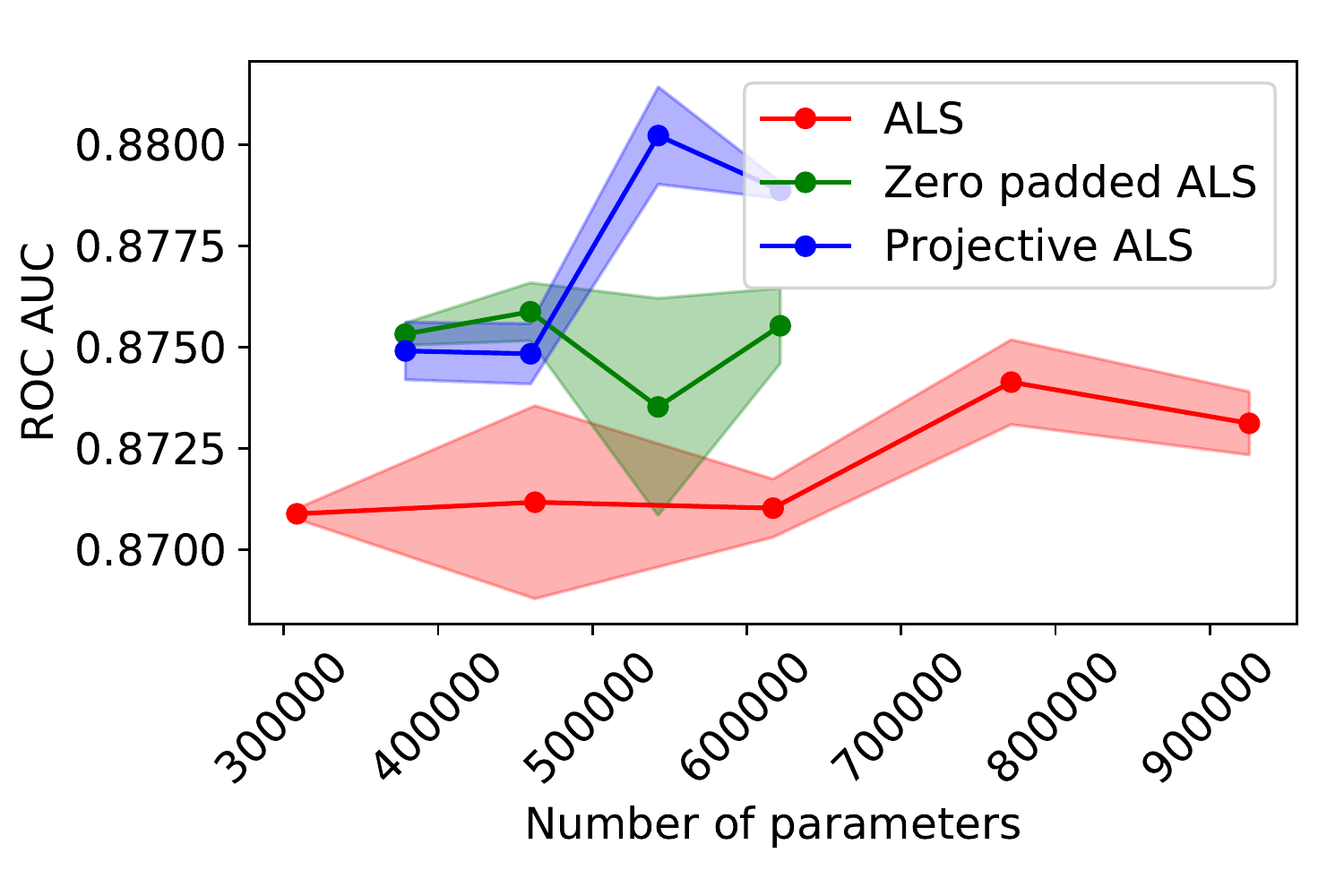}
  \caption{ROC AUC versus the number of parameters on MovieLens 25M dataset.}
  %\Description{The red curve corresponds to ALS, the green curve corresponds to zero padded ALS, the blue curve corresponds to projective ALS. The blue curve is higher than the green curve, the green curve is higher than the red curve.}
  \label{ml25m}
\end{figure}

\begin{figure}[ht]
  \centering
  \includegraphics[width=0.5\linewidth]{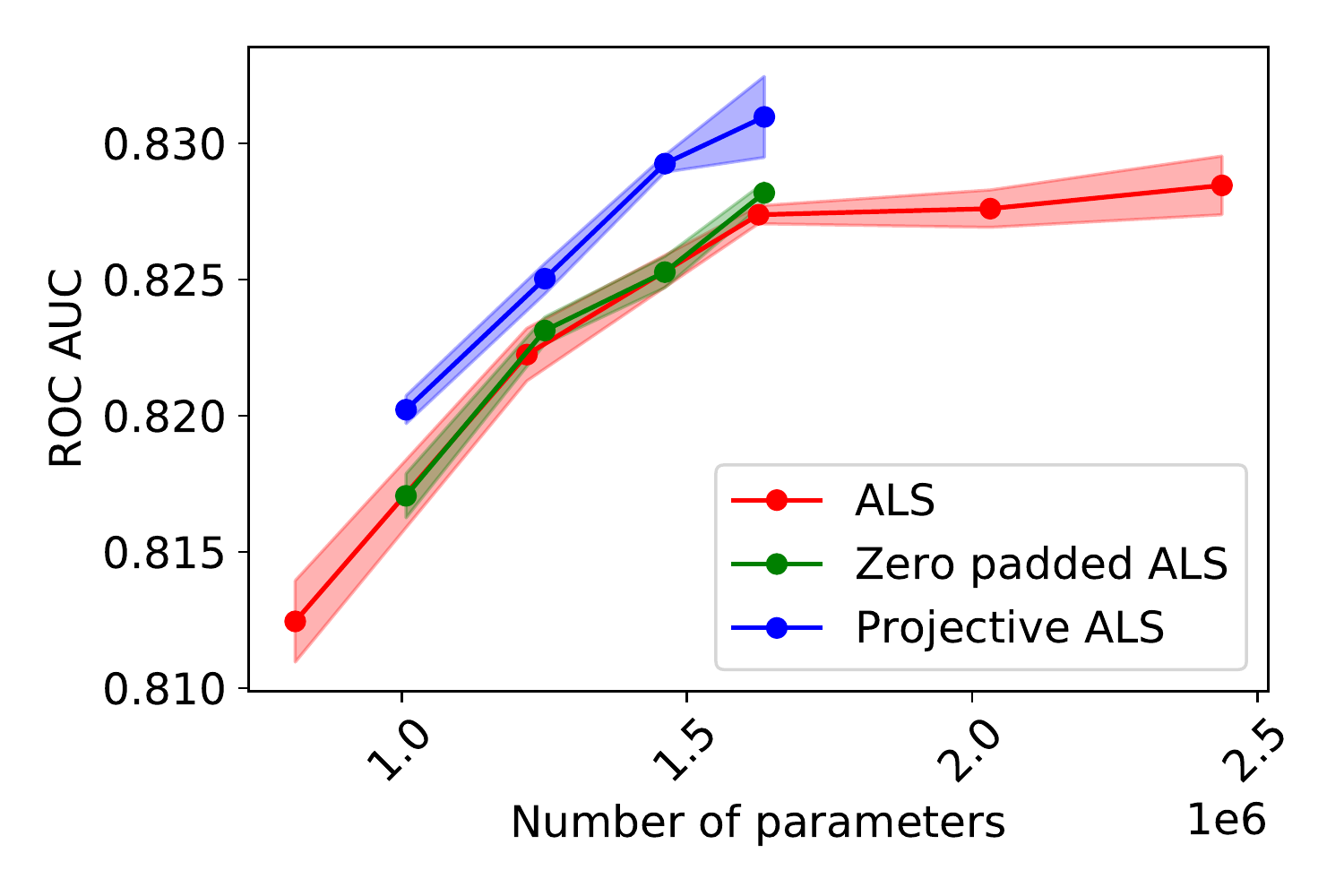}
  \caption{ROC AUC versus the number of parameters on Netflix dataset.}
  %\Description{The red curve corresponds to ALS, the green curve corresponds to zero padded ALS, the blue curve corresponds to projective ALS. The blue curve is higher than the green curve, the green curve matches the red curve.}
  \label{netflix}
\end{figure}

Figures \ref{ml25m} and \ref{netflix} demonstrate the trade-off between ROC AUC and the number of parameters for each of the methods for MovieLens and Netflix datasets respectively. We can see that matrix factorization with projective mixed dimension embeddings exceeds the performance of the baseline for all model sizes on both datasets, and exceeds the performance of the zero padded matrix factorization for most model sizes on both datasets. The zero padded version performs better than the baseline on MovieLens 25M, but only matches it on Netflix dataset.

Comparing the models of the maximum sizes, we can see that in both cases ALS with projective embeddings yields better performance, while using approximately $35\%$ less parameters.

\section{Discussion and Future Work}

Zero padded matrix factorization can be interpreted as a special case of the projective matrix factorization. If we choose all the projection matrices to be rectangular diagonal matrices with diagonal filled with ones and fix them, the model will be equivalent to the zero padded version. So we can conclude that the projective version is more expressive due to a small number of additional parameters: non-diagonal projection matrices allow for interaction between all the embedding components.

In this paper we focus only on mixed dimension embeddings in application to the explicit feedback setting. In this setting, user's preferences are known for a small set of items, and all unknown user-item pairs are ignored. Similar methods are also applicable for implicit feedback setting, where unknown user-item pairs are assumed to have rating 0 and a low weight. We leave this for future work.

\section{Conclusion}

In this paper we propose two methods for matrix factorization with mixed dimension embeddings. Both of the proposed models can be optimized with alternating least squares in a massively parallel manner, which makes them applicable to big data scenario. We analyse the complexity of out methods and show that it is linear in the number of users, items and interactions. 

We evaluate our methods on two popular datasets for collaborative filtering and demonstrate that they offer a better complexity-performance trade-off than the baseline.

%\section*{Acknowledgements}

\bibliographystyle{unsrt}  
\bibliography{main}

\end{document}